\documentclass[pre,preprint, superscriptaddress, amsmath, amssymb, floatfix]{revtex4}
\usepackage{graphicx}%
\usepackage{dcolumn} %
\usepackage{bm}      %
\usepackage{subfigure}
\usepackage{amsfonts}
\usepackage[usenames]{color}
\usepackage{rotating}
\usepackage{fontenc}
\usepackage{cancel}
\usepackage[normalem]{ulem}

\usepackage{amsthm}

\usepackage{hyperref}

\newcommand{\be}{\begin{equation}}
\newcommand{\ee}{\end{equation}}
\newcommand{\bea}{\begin{eqnarray}}
\newcommand{\eea}{\end{eqnarray}}
\newcommand{\mm}{\mathrm}

\newcommand{\bi}{\begin{itemize}}
\newcommand{\ei}{\end{itemize}}

\newcommand{\kmean}{\langle k \rangle}
\newcommand{\kout}{k_\mm{out}}
\newcommand{\kin}{k_\mm{in}}

\begin{document}
\title{Control centrality and hierarchical structure in complex networks}
\author{Yang-Yu Liu}  %
\affiliation{Center for Complex Network Research and Department of
  Physics,Northeastern University, Boston, Massachusetts 02115, USA} 
\affiliation{Center for Cancer Systems Biology, Dana-Farber Cancer Institute, Boston, Massachusetts 02115, USA.}
\author{Jean-Jacques  Slotine}
\affiliation{Nonlinear Systems Laboratory, Massachusetts Institute of Technology, Cambridge, Massachusetts, 02139, USA}
\affiliation{Department of Mechanical Engineering and Department of Brain and
  Cognitive Sciences, Massachusetts Institute of Technology, Cambridge,
  Massachusetts, 02139, USA} 
\author{Albert-L\'{a}szl\'{o} Barab\'{a}si}
\affiliation{Center for Complex Network Research and Department of
  Physics,Northeastern University, Boston, Massachusetts 02115, USA} 
\affiliation{Center for Cancer Systems Biology, Dana-Farber Cancer Institute, Boston, Massachusetts 02115, USA.}
\affiliation{Department of Medicine, Brigham and Women's Hospital, Harvard
  Medical School, Boston, Massachusetts 02115, USA.} 
\date{\today}

\begin{abstract}
  We introduce the concept of control centrality to
  quantify the ability of a single node to control a directed weighted
  network. 
We calculate the distribution of control centrality for several real
networks and find that it is mainly determined by the network's degree
distribution. 
  We rigorously prove that in a directed network without loops 
 the control centrality of a node is uniquely determined by its
  layer index or topological position in the underlying hierarchical
  structure of the network.
  Inspired by the deep relation between control centrality and
  hierarchical structure in a general directed network, we 
  design an efficient attack strategy against the controllability of
  malicious networks.   

\end{abstract}

\maketitle
Complex networks have been at the forefront of statistical mechanics for more than a
decade~\cite{Albert-RMP-02,Newman-Book-06,Barabasi-Science-99,Watts-Nature-98}. 
Studies of them impact our understanding and control of a wide range
of systems, from Internet and the power-grid to cellular and
ecological networks. 
Despite the diversity of complex networks, several basic universal
principles have been uncovered that govern their
topology and evolution~\cite{Barabasi-Science-99,Watts-Nature-98}. 
While these principles have significantly enriched our understanding
of many networks that affect our lives, our ultimate goal
is to develop the capability to control
them~\cite{Wang-PA-02,Tanner-IEEE-04, Sorrentino-PRE-07,Yu-Automatica-09,
    Lombardi-PRE-07, Rahmani-SIAM-09, Mesbahi-Book-10,
    Liu-Nature-11,Liu-NatureReply-11,Egerstedt-Nature-11,
    Nepusz-arXiv-11,Cowan-arXiv-11,Wang-PRE-12}.

According to control theory, a dynamical system is controllable
if, with a suitable choice of inputs, it can be driven from any
initial state to any desired final state in finite
time~\cite{Kalman-JSIAM-63,Luenberger-Book-79,Slotine-Book-91}. 
By combining tools from control theory and network science, we
proposed an efficient methodology to identify the minimum sets of
driver nodes, whose time-dependent control can guide the whole network
to any desired final state~\cite{Liu-Nature-11}. %
Yet, this minimum driver set (MDS) is usually
not unique, but one can often achieve multiple potential control configurations
with the same number of driver nodes. Given that some nodes may appear
in some MDSs but not in other, a crucial question remains unanswered:
what is the role of each individual node in controlling a complex system? 
Therefore the question that we address in this paper pertains to
the importance of a given node in maintaining a system's controllability.  
Consider a complex system described by a directed weighted network of
$N$ nodes whose time evolution follows the linear time-invariant
dynamics  
\be
\dot{\bf x}(t) = {\bf A} \, {\bf x}(t) + {\bf B}\, {\bf u}(t)
\label{eq:X}\ee 
where ${\bf x}(t)=(x_1(t),x_2(t),\cdots,x_N(t))^\mm{T} \in \mathbb{R}^N$ captures the
state of each node at time $t$. 
${\bf A} \in \mathbb{R}^{N \times N}$ is an $N \times N$ matrix describing the weighted wiring
diagram of the network. The matrix element $a_{ij} \in \mathbb{R}$ gives the
strength or weight that node $j$ can affect node $i$. Positive (or
negative) value of $a_{ij}$ means the link $(j \to i)$ is excitatory
(or inhibitory). 
${\bf B} \in \mathbb{R}^{N \times M}$ is an $N \times M$ input matrix ($M\le N$) identifying the
nodes that are controlled by the time dependent input vector ${\bf
  u}(t) = (u_1(t),u_2(t),\cdots,u_M(t))^\mm{T} \in \mathbb{R}^{M}$ with $M$ independent
signals %
imposed by an outside controller. 
The matrix element $b_{ij} \in \mathbb{R}$ represents the coupling
strength between the input signal $u_j(t)$ and node $i$. 
The system (\ref{eq:X}), also denoted as $({\bf A}, {\bf B})$, is
controllable if and only if its controllability matrix ${\bf C}=({\bf B}, {\bf AB}, \cdots, {\bf
  A}^{N-1}{\bf B}) \in \mathbb{R}^{N\times
NM}$ has full rank, a criteria often called Kalman's
controllability rank condition~\cite{Kalman-JSIAM-63}. 
The rank of the controllability matrix ${\bf C}$, denoted by
$\mm{rank}({\bf C})$, provides the dimension of the controllable subspace of the
system $({\bf A}, {\bf B})$~\cite{Kalman-JSIAM-63,Luenberger-Book-79}. 
When we control node $i$ only, %
${\bf B}$ reduces to the vector ${\bf
  b}^{(i)}$ with a single non-zero entry,  and we denote ${\bf C}$
with ${\bf C}^{(i)}$. 
We can therefore use $\mm{rank}({\bf C}^{(i)})$ as a natural measure of
node $i$'s ability to control the system:  
if $\mm{rank}({\bf C}^{(i)})=N$, then node $i$ alone can control the whole
system, i.e. it can drive the system between any points in the $N$-dimensional
state space in finite time. Any value of $\mm{rank}({\bf C}^{(i)})$ less than $N$ provides the
dimension of the subspace $i$ can control. In particular if $\mm{rank}({\bf C}^{(i)})=1$,
then node $i$ can only control itself.

\begin{figure}[t!]
\includegraphics[width=0.8\textwidth]{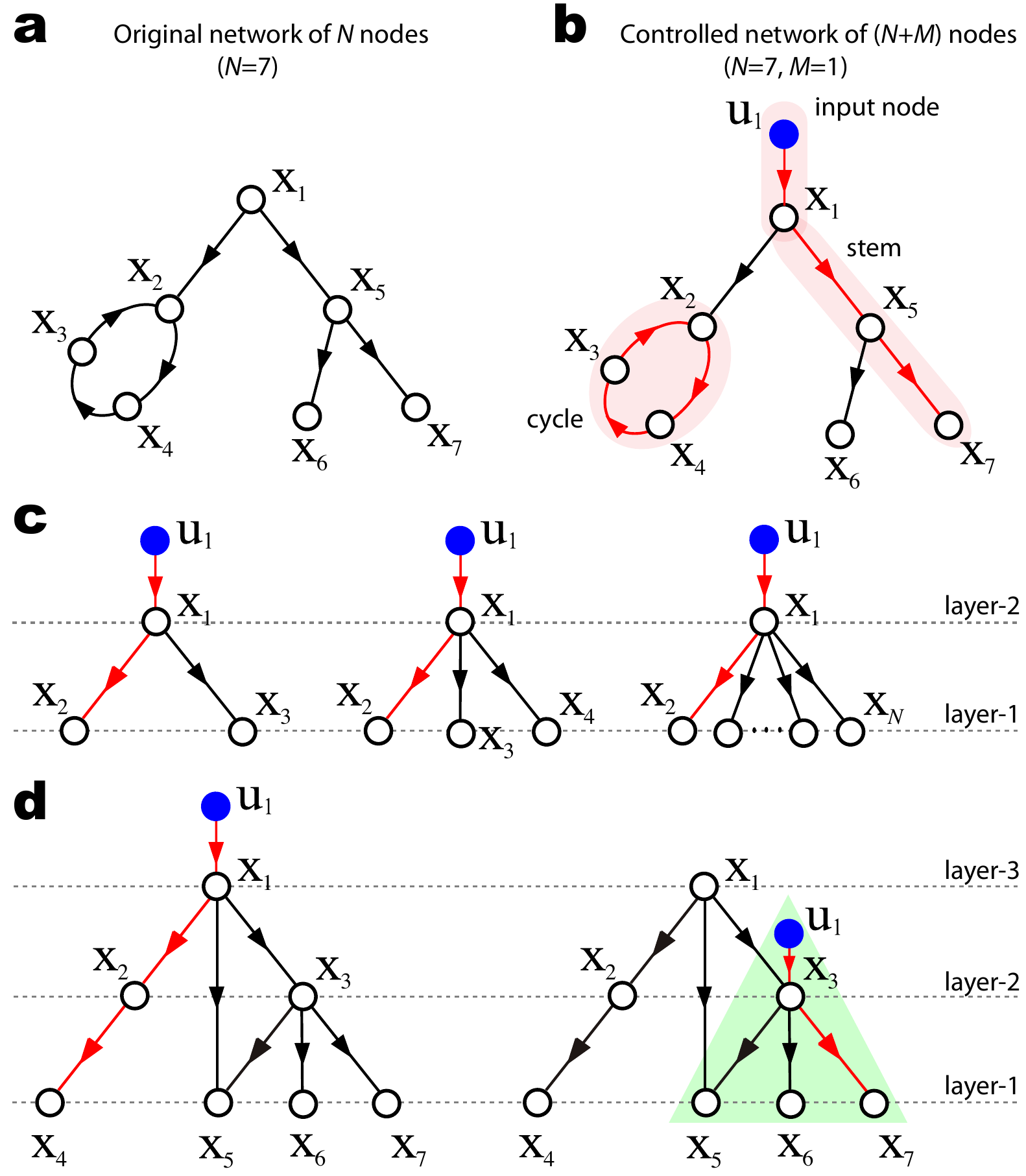}
  \caption{\label{fig:Hosoe} {\bf Control centrality.} (a) A simple
    network of $N=7$ nodes. (b) The controlled network is represented by a directed graph
    $G({\bf A}, {\bf B})$ with an input node $u_1$ connecting
    to a state node $x_1$. 
    The stem-cycle disjoint subgraph $G_\mm{s}$ (shown in red)
    contains six edges, which is the largest number of edges among all
    possible stem-cycle disjoint subgraphs of the directed graph $G({\bf A},
    {\bf B})$ and corresponds to the generic dimension of
    controllable subspace by controlling node $x_1$. The control
    centrality of node $1$ is thus $C_\mm{c}(1)=6$.
(c) The control centrality of the central hub in a directed star is
always 2 for any network size $N\ge 2$.
(d) The control centrality of a node in a directed acyclic graph (DAG)
equals its layer index. In applying Hosoe's theorem, if not all state
nodes are accessible, we just need to consider the accessible part
(highlighted in green) of the input node(s).} 
\end{figure}

The precise value of $\mm{rank}({\bf C})$ is difficult to %
determine because in reality the system parameters, i.e. the elements
of ${\bf A}$ and ${\bf B}$, are often not known precisely except the
zeros that mark the absence of connections between components of the
system~\cite{Lin-IEEE-74}. 
Hence ${\bf A}$ and ${\bf B}$ are often considered to be structured matrices, i.e. their elements
are either fixed zeros or independent free parameters~\cite{Lin-IEEE-74}. 
Apparently, $\mm{rank}({\bf C})$ varies as a function of the free
parameters of ${\bf A}$ and ${\bf B}$. However, it achieves the
maximal value for all but an exceptional set of values of the free
parameters which forms a proper variety with Lebesgue measure zero in
the parameter space~\cite{Shields-IEEE-76,Hosoe-IEEE-80}.  
This maximal value is called the \emph{generic rank} of the
controllability matrix ${\bf C}$, denoted as $\mm{rank}_\mm{g}({\bf
  C})$, which also represents the generic dimension of the
controllable subspace.  
 When $\mm{rank}_\mm{g}({\bf
  C}) = N$, the system $({\bf A}, {\bf B})$ is \emph{structurally
  controllable}, i.e. 
controllable for almost all sets of values of
 the free parameters of ${\bf A}$ and ${\bf B}$ 
except an exceptional set of values with zero
measure~\cite{Lin-IEEE-74,Shields-IEEE-76,Dion-Automatica-03,Blackhall-IFAC-10}.
For a single node $i$, $\mm{rank}_\mm{g}({\bf C}^{(i)})$ captures the ``power'' of $i$ in controlling the
whole network, allowing us to define the \emph{control centrality} of node $i$ as 
\be
C_\mm{c}(i) \equiv \mm{rank}_\mm{g}({\bf C}^{(i)}).
\ee %
The calculation of $\mm{rank}_\mm{g}({\bf C})$ can be mapped into a combinatorial 
optimization problem on a directed graph $G({\bf A},{\bf B})$ constructed
as follows~\cite{Hosoe-IEEE-80}. 
Connect the $M$ input nodes $\{u_1,\cdots, u_M\}$ to the $N$ state
nodes $\{x_1,\cdots, x_N\}$ in the original network according to the
input matrix ${\bf B}$, i.e. connect $u_j$ to $x_i$ if $b_{ij} \neq
0$, obtaining a directed graph $G({\bf A},{\bf B})$ with
$N+M$ nodes (see Fig.~\ref{fig:Hosoe}a and b). 
A state node $j$ is called \emph{accessible} if there is at least one
directed path reaching from one of the input nodes to node $j$. 
In Fig.~\ref{fig:Hosoe}b, all state nodes $\{x_1,\cdots, x_7\}$ are
accessible from the input node $u_1$. 
A \emph{stem} is a directed path starting from an input node, so that no 
nodes appear more than once in it, e.g. $u_1 \to x_1 \to x_5 \to x_7$
in Fig.~\ref{fig:Hosoe}b. 
Denote with $G_\mm{s}$ the \emph{stem-cycle disjoint} subgraph of $G({\bf A}, {\bf
  B})$, such that
$G_\mm{s}$ consists of stems and cycles only, and the stems and cycles
have no node in common (highlighted in Fig.~\ref{fig:Hosoe}b). 
According to Hosoe's theorem\cite{Hosoe-IEEE-80},  the generic
dimension of the controllable subspace is given by  
\be \mm{rank}_\mm{g}({\bf C}) = \max_{G_\mm{s} \in {\cal G}} {|E(G_\mm{s})|} \label{eq:EG}
\ee 
with ${\cal G}$ the set of all stem-cycle disjoint subgraphs of the
accessible part of $G({\bf A}, {\bf B})$ and $|E(G_\mm{s})|$ the number of edges in the subgraph
$G_\mm{s}$. 
For example, the subgraph highlighted in Fig.~\ref{fig:Hosoe}b,
denoted as $G_\mm{s}^\mm{max}$, contains the largest number of edges
among all possible stem-cycle disjoint subgraphs.  
Thus, $C_\mm{c}(1) = \mm{rank}_\mm{g}({\bf
  C}^{(1)}) = 6$, which is the number of red links in Fig.~\ref{fig:Hosoe}b. 
Note that $\mm{rank}_\mm{g}({\bf C}^{(1)}) = 6 < N =7$, the whole system
is therefore not structurally controllable by controlling $x_1$
only. Yet, the nodes covered by the $G_\mm{s}^\mm{max}$ highlighted in
Fig.~\ref{fig:Hosoe}b, e.g. $\{x_1, x_2, x_3, x_4, x_5, x_7\}$,
constitute a structurally controllable subsystem~\cite{Blackhall-IFAC-10}. 
In other words, by controlling node $x_1$ with a time dependent signal
$u_1(t)$ we can drive the subsystem $\{x_1, x_2, x_3, x_4, x_5,
x_7\}$ from any initial state to any final state in finite time, for
almost all sets of values of the free parameters of ${\bf A}$ and
${\bf B}$ except an exceptional set of values with zero measure. 
In general $G_\mm{s}^\mm{max}$ is not unique. For example, in
Fig.~\ref{fig:Hosoe}b we can get the same cycle $x_2 \to x_3 \to x_4
\to x_2$ together with a different stem $u_1 \to x_1 \to x_5 \to
x_6$, which yield a different $G_\mm{s}^\mm{max}$ and thus a different
structurally controllable subsystem $\{x_1, x_2, x_3, x_4, x_5, x_6\}$. 
Both subsystems are of size six, which is exactly the generic
dimension of the controllable subspace. Note that we can fully control
each subsystem individually, yet we cannot fully control the whole
system.

The advantage of Eq.(\ref{eq:EG}) is that $\max_{G_\mm{s} \in {\cal
    G}} {|E(G)|}$ can be calculated via linear
programming~\cite{Poljak-IEEE-90}, providing us an efficient numerical
tool to determine the control centrality and the structurally
controllable subsystem of any node in an arbitrary complex network
(see Supplementary Material Sec.I.A).

\begin{figure}[t!]
\includegraphics[width=\textwidth]{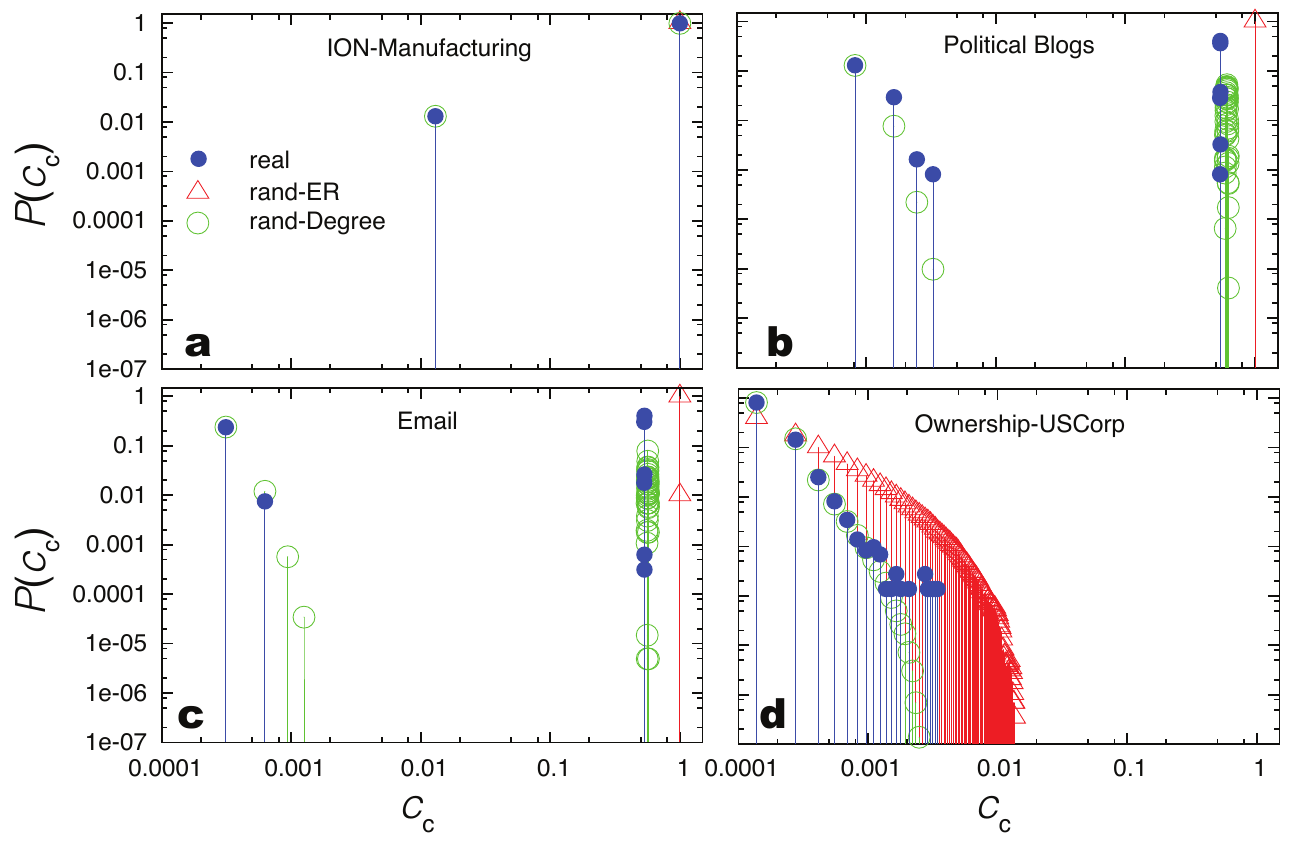}
 \caption{\label{fig:PCs} {\bf Distribution of normalized control
     centrality of several real-world networks (blue) and their 
     randomized counterparts: rand-ER (red),
     rand-Degree (green).} 
(a) Intra-organizational network of a manufacturing
   company~\cite{Parker-Book-04}.
(b) Hyperlinks between weblogs on US politics~\cite{Adamic-05}.  
(c) Email network in a university~\cite{Eckmann-PNAS-04}. 
(d) Ownership network of US corporations~\cite{Norlen-02}. 
}
\end{figure}

We first consider the distribution of control centrality. Shown in
Fig.~\ref{fig:PCs} is the distribution of the normalized control
centrality ($c_\mm{c}(i) \equiv C_\mm{c}(i)/N$) for several real networks. 
We find that for the intra-organization network, $P(c_\mm{c})$ has a sharp peak at
$c_\mm{c}=1$, suggesting that a high fraction of nodes can individually exert
full control over the whole system (Fig.~\ref{fig:PCs}a). In contrast, for
company-ownership network, $P(c_\mm{c})$ follows an approximately exponential
distribution (Fig.~\ref{fig:PCs}d), indicating that most nodes display low
control centrality. Even the most powerful node, %
with $c_\mm{c} \sim 0.01$, %
can control only one percent
of the total dimension of the system's full state space. 
For other networks $P(c_\mm{c})$ displays a mixed behavior, indicating
the coexistence of a few powerful nodes with a large number of nodes
that have little control over the system's dynamics (Fig.~\ref{fig:PCs}b,c). 
Note that under full randomization, turning a network into a directed
Erd\H{o}s-R\'enyi (ER) random
network~\cite{Erdos-PMIHAS-60,Bollobas-Book-01} with number of nodes
($N$) and number of edges ($L$) unchanged, the $c_\mm{c}$ distribution changes dramatically. 
In contrast, under degree-preserving randomization
~\cite{Maslov-Science-02,Milo-Science-02}, which keeps the in-degree
($k_\mm{in}$)  and out-degree ($k_\mm{out}$)  of each node unchanged, 
the $c_\mm{c}$ distribution does not change significantly. 
This result suggests that $P(c_\mm{c})$ is mainly determined by the underlying 
network's degree distribution $P(\kin, \kout)$. 
This result is very useful in the following sense:
$P(\kin, \kout)$ is easy to calculate for any complex network, while the
calculation of $P(c_\mm{c})$ requires much more computational efforts (both
CPU time and memory space). Studying $P(c_\mm{c})$ for model networks
of prescribed $P(\kin, \kout)$ will give us qualitative understanding
of how $P(c_\mm{c})$ changes as we vary network parameters, e.g. mean
degree $\kmean$. See Supplementary Material Sec.II for more
details. 
To understand which topological features determine the control centrality
itself, we compared the control centrality for each node in the real
networks 
and their randomized counterparts (denoted as rand-ER and rand-Degree). %
The lack of correlations indicates that both 
randomization procedures eliminate the topological feature that determines the
control centrality of a given node (see Supplementary Material Sec.I.B). 
Since accessibility plays an important role in maintaining structural
controllability~\cite{Lin-IEEE-74}, we conjecture that the control centrality
of node $i$ is correlated with the number of nodes $N_\mm{r}(i)$ that can be
reached from it. To test this conjecture, we calculated $N_\mm{r}(i)$ and $C_\mm{c}(i)$
for the real networks shown in Fig.~\ref{fig:PCs}, observing only a weak
correlation between the two quantities (see Supplementary
Material Sec.I.C). This lack of correlation between $N_\mm{r}(i)$ and
$C_\mm{c}(i)$ is obvious in a directed star, in which a central hub ($x_1$)
points to $N-1$ leaf nodes ($x_2, \cdots, x_{N}$) (Fig.~\ref{fig:Hosoe}c).  As
the central hub can reach all nodes, $N_\mm{r}(1)=N$, suggesting that
it should have high control centrality. Yet, one can easily
check that the central hub has control centrality
$C_\mm{c}(1) = 2$ for any $N\ge 2$ and there are $N-1$ structurally
controllable subsystems, i.e. $\{x_1, x_2\}, \cdots,
\{x_1,x_{N-1}\}$. In other words, by controlling the central hub we
can fully control each leaf node individually, but we cannot control
them collectively. 

Note that in a directed star each node can
be labeled with a unique \emph{layer index}: the leaf nodes are in the
first layer (bottom layer) and the central hub is in the 
second layer (top layer). In this case the control centrality of the
central hub equals its layer index (see Fig.~\ref{fig:Hosoe}c). This is not by coincidence: 
we can prove that for a directed network containing no cycles,  often
called a directed acyclic graph (DAG), the control centrality of any node equals its \emph{layer index} 
  \be C_\mm{c}(i) = l_i .\label{eq:CL}\ee
Indeed, lacking cycles, a DAG has a unique \emph{hierarchical structure}, which
means that each node can be labeled with a unique layer index ($l_i$),
calculated using a recursive labeling algorithm~\cite{Yan-PNAS-10}: (1) Nodes
that have no outgoing links ($k_\mm{out}=0$) are labeled with layer index 1
(bottom layer). (2) Remove all nodes in layer 1. For the remaining
graph identify again all nodes with $k_\mm{out}=0$ and label them with
layer index 2. (3) Repeat step (2) until all nodes are labeled. 
As the DAG lacks cycles, each subgraph in the set ${\cal
  G}$ of the directed graph $G({\bf A}, {\bf b}^{(i)})$ consists of a stem
only, which starts from the input node pointing to the state node $i$
and ends at a state node
in the bottom layer, e.g. $u_1 \to x_1 \to x_2 \to x_4$ in Fig.~\ref{fig:Hosoe}d. The number of edges in this stem is equal to the layer
index of node $i$, so  $\mm{rank}_\mm{g}({\bf C}^{(i)})=C_\mm{c}(i) =
l_i$. %
Therefore in DAG the higher a node is in the hierarchy, the higher is its
ability to control the system. Though this result agrees with our
intuition to some extent, it is surprising at the first glance because
it indicates that in a DAG the control centrality of node $i$ is only determined
by its topological position in the hierarchical structure, rather than
any other importance measures, e.g. degree or betweenness centrality.  
This result also partially explains why driver nodes tend to avoid
hubs~\cite{Liu-Nature-11}. 
Despite the simplicity of Eq.~(\ref{eq:CL}), we cannot apply it directly to real
networks, because most of them are not DAGs. Yet, we note that any
directed network has a underlying DAG structure based on the strongly
connected component (SCC) decomposition (see Supplementary Material
Sec.I.D Fig.~S4).  
A subgraph of a directed network is \emph{strongly connected} if there is
a directed path from each node in the subgraph to every other node. 
The SCCs of a directed network $G$ are its maximal strongly connected
subgraphs.  
If we contract each SCC to a single supernode,
the resulting graph $\widetilde{G}$, called the \emph{condensation}
of $G$, is a DAG~\cite{Harary-Book-94}.    
Since a DAG has a unique hierarchical structure, a directed
network can then be assigned an underlying hierarchical 
structure. The layer index of node $i$ can be defined to be the layer
index of the corresponding supernode (i.e. the SCC
that node $i$ belongs to) in $\widetilde{G}$. With this definition of
$l_i$, it is easy to show that $C_\mm{c}(i) \ge l_i$ for general
directed networks. 
Furthermore, for an edge $(i\to j)$ in a general
directed network, if node $i$ is topologically ``higher'' than node
$j$ (i.e. $l_i > l_j$), then $C_\mm{c}(i) >
C_\mm{c}(j)$.   
Since $C_\mm{c}(i)$ has to be calculated via linear programming which is
computationally more challenging %
than the calculation of $l_i$, the above results suggest an efficient
way to calculate the lower bound of $C_\mm{c}(i)$ and to compare the
control centralities of two neighboring nodes. 
Note that if $l_i > l_j$ and there is no directed edge $(i \to j)$ in
the network, then in general one cannot conclude
  that $C_\mm{c}(i) >
C_\mm{c}(j)$ (see Supplementary Material
Sec.I.D for more details). 
Our finding on the relation between control centrality and
hierarchical structure inspires us to design an efficient attack
strategy against malicious networks, aiming to affect their
controllability.   
The most efficient way to damage the controllability of a network is
to remove all input nodes $\{u_1, u_2, \cdots, u_M\}$, rendering the
system completely uncontrollable. But this requires a detailed
knowledge of the control configuration, i.e. the wiring diagram of
$G({\bf A}, {\bf B})$, which we often lack.
If the network structure (${\bf A}$) is known, one can attempt a
\emph{targeted attack}, i.e. rank the nodes according to some
centrality measure, like degree or control centrality, and remove the
nodes with highest centralities~\cite{Albert-Nature-00, Cohen-PRL-03}.  
Though we still lack systematic studies on the effect of a targeted
attack on a network's controllability, one naively expects that this 
should be the most efficient strategy. 
But we often lack the knowledge of the network
structure, which makes this approach unfeasible anyway. 
In this case a simple strategy would be \emph{random
  attack}, i.e. remove a randomly chosen $P$ fraction of nodes, which
naturally serves as a benchmark for any other strategy.  
Here we propose instead a \emph{random
  upstream attack} strategy: randomly choose a $P$ fraction of nodes, and for
each node remove one of its incoming or upstream neighbors if it has
one, otherwise remove the 
node itself. A \emph{random downstream} attack can be defined
similarly, removing the node to which the chosen node
points to. 
In undirected networks, a similar strategy has been proposed
for efficient immunization~\cite{Cohen-PRL-03} and the early detection
of contagious outbreaks~\cite{Christakis-PLoS-10}, relying on
the statistical trend that %
randomly selected neighbors have more
links than the node itself~\cite{Feld-AJS-91,Newman-SN-03}. In
directed networks we can prove that 
randomly selected upstream (or downstream) neighbors have more
outgoing (or incoming) links than the node itself (see Supplementary
Material Sec.III.A). Thus a random upstream (or downstream) attack will
remove more hubs and more links than the random
attack does. But the real reason why we expect a random upstream
attack to be efficient in a directed network is because $C_\mm{c}(i) \ge
C_\mm{c}(j)$ for most edges $(i\to j)$, i.e. the control centrality of the starting
node is usually no less than the ending node of a directed edge. In
DAGs, for any edge $(i\to j)$, we have strictly $C_\mm{c}(i) >
C_\mm{c}(j)$ (see
Supplementary Material Sec.III.B). Thus, %
the upstream neighbor of a node is expected to play a more
important or equal role in control than the node itself, a result
deeply rooted in the nature of the control problem, rather than the
hub status of the upstream nodes.   

\begin{figure}
\includegraphics[width=\textwidth]{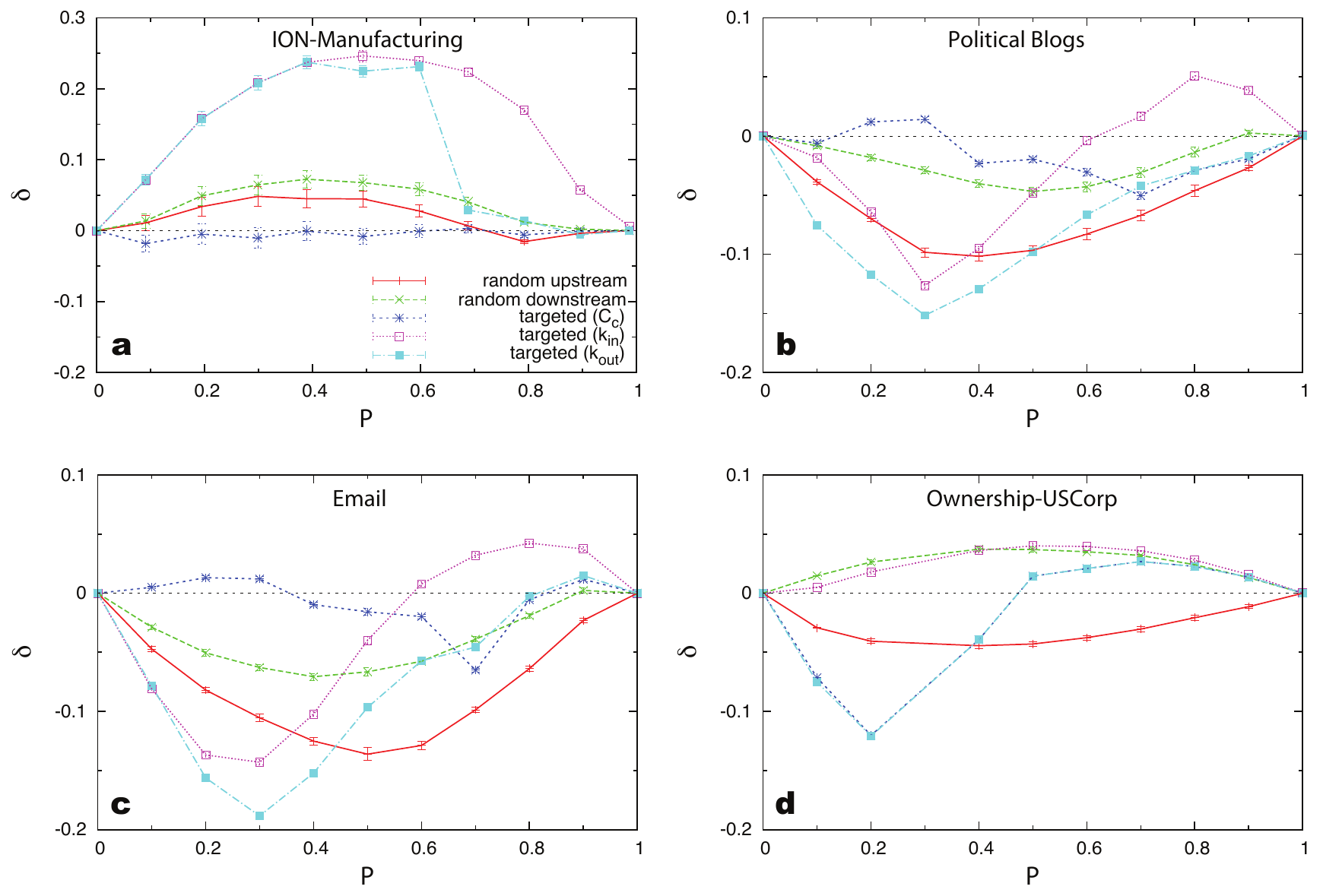}
\caption{\label{fig:attack-P} {\bf The impact of different attack
     strategies on network controllability with respective to the random attack.} $\delta \equiv 
   [\mm{rank}_\mm{g}^{\mm{Strategy}-j}({\bf C}') -
   \mm{rank}_\mm{g}^\mm{Random}({\bf C}')]/N$ with $\mm{rank}_\mm{g}^{\mm{Strategy}-j} ({\bf C}')$ represents the generic dimension of controllable
    subspace after removing a $P$ fraction of nodes using strategy-$j$.  The nodes are
  removed according to six different strategies. (Strategy-0) Random attack:
  randomly remove $P$ fraction of nodes. (Strategy-1 or 2) Random
  upstream (or downstream) attack: randomly choose $P$ fraction of nodes, randomly
  remove one of their upstream neighbors (or downstream neighbors). The results are
    averaged over 10 random choices of $P$ fraction of nodes with
    error bars defined as s.e.m. Lines are only a guide to the eye.  (Strategy-3,4, or 5) Targeted
  attacks: remove the top $P$
  fraction of nodes according to their control centralities (or
  in-degrees or out-degrees). 
}
\end{figure}

To show the efficiency %
of the random upstream attack we compare
its impact on fully controlled networks with several other 
strategies.   
We start from a network that is fully controlled ($\mm{rank}_\mm{g}({\bf
  C})=N$) via a minimum set of 
$N_\mm{D}$ driver nodes. After the attack a $P$ faction of
nodes are removed, denoting with $\mm{rank}_\mm{g}({\bf C}')$ the dimension
of the controllable subspace of the damaged network. We calculate $\mm{rank}_\mm{g}({\bf C}')$
as a function of $P$, with $P$ tuned from 0 up to 1. 
Since the random attack %
serves as a natural benchmark, we
calculate the difference of $\mm{rank}_\mm{g}({\bf C}')$ between a given strategy and the
random attack, denoted as $\delta = [\mm{rank}_\mm{g}^{\mm{Strategy}-j}({\bf
  C}') - \mm{rank}_\mm{g}^{\mm{Random}} ({\bf C}')]/N$. 
Apparently, the more negative is $\delta$, the more efficient is the
strategy compared to a fully random attack. 
We find that for most networks random upstream attack
results in $\delta<0$ for $0<P<1$, i.e. it causes more damage to
the network's controllability than random
attack (see Fig.~\ref{fig:attack-P}b,c,d). 
Moreover, random upstream attack typically is more efficient than
random downstream attack, even though in both cases we remove more
hubs and more links than in the random attack. This is due to the
fact that the upstream (or downstream) neighbors are usually more (or
less) ``powerful'' than the node itself.  
The efficiency %
of the random upstream attack is even comparable to
targeted attacks (Fig.~\ref{fig:attack-P}). Since the former requires
only the knowledge of the network's local structure rather than any knowledge of the
nodes' centrality measures or any other global information (i.e. the
structure of the ${\bf A}$ matrix) while the latter rely heavily on
them, this finding indicates the advantage of the random upstream
attack. 
The fact that those targeted attacks do not always show significant
superiority over the random attack and the random upstream (downstream)
attack could be due to an \emph{overlap effect} --- two targeted
nodes successively chosen from the rank list based on some centrality
measure are likely to have larger overlap between their controllable
subspaces than two randomly chosen nodes. 
Therefore, successively removing targeted nodes with highest
centralities will not always cause the most damage to the network's controllability.  
Random attacks can avoid such overlap to some extent because the
removed nodes are randomly chosen. 
This also explains why sometimes targeted attacks are worse than the
random attack (see Fig.~\ref{fig:attack-P}a). 
Notice that for the intra-organization network all attack strategies
fail in the sense that $\delta$ is either positive or very close to
zero (Fig.~\ref{fig:attack-P}a). This is due to the fact
this network is so dense ($\kmean \approx 58$) that we have
$C_\mm{c}(i) = C_\mm{c}(j) =N$ for almost all the edges $(i\to j)$. Consequently,
both random upstream and downstream attacks are not efficient and the
$C_\mm{c}$-targeted attack shows almost the same impact as the random
attack. 
This result suggests that when the network becomes very dense its
controllability becomes extremely robust against all kinds of attacks,
consistent with our previous result on the core percolation and the
control robustness against link removal~\cite{Liu-Nature-11}. 
We also tested those attack strategies on model networks (see
Supplementary Material Sec.III.C). The results are qualitatively consistent with
what we observed in real networks. 
In sum, we study the control centrality of single node in complex
networks and find that it is related to the underlying
hierarchical structure of networks. 
The presented results help us better understand the 
controllability of complex networks and design an efficient attack
strategy against network control. 
This work was supported by the Network
Science Collaborative Technology Alliance sponsored by the US Army
Research Laboratory under Agreement Number W911NF-09-2-0053; the
Office of Naval Research under Agreement Number N000141010968; the
Defense Threat Reduction Agency awards WMD BRBAA07-J-2-0035 and
BRBAA08-Per4-C-2-0033; and the James S. McDonnell Foundation 21st
Century Initiative in Studying Complex Systems.

\newpage

\end{document}